\documentclass[aps,prd,twocolumn,nofootinbib]{revtex4-2}

\usepackage{graphicx,epsfig}
\usepackage{amsmath,amssymb,mathrsfs,slashed,bm}
\usepackage{amsfonts}
\usepackage{float}

\allowdisplaybreaks[4]

\usepackage[colorlinks=true,linkcolor=blue,citecolor=blue, urlcolor=blue]{hyperref}

\newcommand{\Rti}{R_{\tau}}
\newcommand{\as}{\alpha_s}

\newcommand{\err}[3]{#1^{+#2}_{-#3}}

\begin{document}

\title{A new method for estimating unknown one-order higher QCD corrections to the perturbative series using the linear regression through the origin}

\author{Zhi-Fei Wu}
\email{wuzf@cqu.edu.cn}
\address{Department of Physics, Chongqing Key Laboratory for Strongly Coupled Physics, Chongqing University, Chongqing 401331, P.R. China}

\author{Xing-Gang Wu}
\email{wuxg@cqu.edu.cn}
\address{Department of Physics, Chongqing Key Laboratory for Strongly Coupled Physics, Chongqing University, Chongqing 401331, P.R. China}

\author{Jiang Yan}
\email{yjiang@itp.ac.cn}
\address{Institute of Theoretical Physics, Chinese Academy of Sciences, Beijing 100190,P.R. China}

\author{Xu-Dong Huang}
\email{huangxd@cqnu.edu.cn}
\address{ College of Physics and Electronic Engineering, Chongqing Normal University, Chongqing 401331, P.R. China}

\author{Jian-Ming Shen}
\email{shenjm@hnu.edu.cn}
\address{School of Physics and Electronics, Hunan Provincial Key Laboratory of High-Energy Scale Physics and Applications, Hunan University, Changsha 410082,P.R. China}

\begin{abstract}

{Quantum Chromodynamics (QCD) is the fundamental theory describing strong interactions. Owing to asymptotic freedom at short distances, high-energy physical observables can be predicted using perturbative QCD (pQCD) following proper factorization. It has been demonstrated that the conventional renormalization scheme-and-scale ambiguities that appear in fixed-order pQCD series can be eliminated by recursively applying the renormalization group equation, aided by the Principle of Maximum Conformality (PMC). To extend the predictive power of pQCD, we still face the challenge of reliably estimating contributions from unknown higher-order (UHO) terms. In this paper, we propose a novel method for estimating one-order higher QCD corrections to the perturbative series: using linear regression through the origin (LRTO) to determine the asymptotic form of the pQCD series below the optimal truncation order $N^*$. When the given $\alpha_s$-order is below $N^*$, its perturbative behavior will be dominated by the usual $\alpha_s$-power suppression and the sub-leading corrections are treated as a source of theoretical uncertainty. This approach enables a quantitative assessment of the series convergence and derives estimate for unknown higher-order contributions. To illustrate this method, we apply it to the important ratio $R_\tau$ which has been calculated up to four-loop QCD corrections. Our results show that the LRTO method yields reliable estimates of the UHO terms, demonstrating its own reliability and significant predictive power for such estimations. In particular, we find that the scale-invariant, more rapidly convergent PMC series exhibits better predictive power -- along with greater stability and reliability -- compared to the initial scale-dependent pQCD series.} 

\end{abstract}
	
\maketitle

\section{Introduction}	\label{intro}

Owing to the property of asymptotic freedom~\cite{Gross:1973id, Politzer:1973fx}, the strong coupling constant \(\alpha_s\) is of order \({\cal O}(1)\) at short distances, and physical observables at high momentum transfers can be systematically computed and expanded as a power series in \(\alpha_s\). The number of terms included in this expansion significantly impacts the accuracy of the perturbative Quantum Chromodynamics (pQCD) predictions. At present, many progresses on the loop calculation technology have been achieved in the literature. However owing to the complexity and the long time-consuming of multi-loop calculations, the pQCD approximations are generally available only at several fixed orders; particularly for the high-energy processes involving hadrons, where renormalization and factorization effects become intricately interconnected and we also have to fix the running behavior of non-pertubative parameters such as the parton distribution functions, the matrix elements and etc. at the required orders. 

On the other hand, the increasing precision of the experimental measurements at high-energy colliders like the LHC underscores the need for more accurate pQCD predictions. Therefore, it is imperative to devise an effective approach for handling the given pQCD series, which not only yields precise fixed-order predictions but also provides a reliable basis for estimating the contributions from unknown higher-order (UHO) terms.

Generally, the pQCD approximant of a physical observable $\rho$ known up to $n_{\rm th}$-order QCD corrections can be expressed as
\begin{align}
	\rho_{n}(\mu_r)=\sum_{k=0}^{n}C_k\as^{k+l}(\mu_r)+{\cal O}(\alpha_s^{n+1+l}),
	\label{generalpQCD}
\end{align}
where $\mu_r$ is the renormalization scale, $l$ denotes the $\alpha_s$-power at tree-level ($n=0$) or leading-order (LO) level. The uncertainties of this initial fixed-order series (\ref{generalpQCD}) are usually separated into two parts. One of these is the scale uncertainty, which is conventionally estimated by varying the renormalization scale $\mu_r$ within a specific range -- for example, $\mu_r\in[Q/2, 2Q]$, where $Q$ denotes typical momentum flow of the process. However, this naive way only roughly estimates the uncertainty from different choices of $\mu_r$ and partly estimates the $\beta$-dependent non-conformal UHO contributions that control the running behavior of $\alpha_s$. Thus, it can only serve as a qualitative order-of-magnitude estimation and cannot function as the desired precise quantitative estimation of the UHO-terms. Then, the question arises: ``How to achieve a fair estimation of the magnitudes of the higher-order terms in the perturbation series without having to laboriously calculate Feynman diagrams ? ". This question was already raised by Feynman in 1961~\cite{Feynman}. To date, three typical types of approaches have been proposed for estimating the contribution of the UHO-terms:
\begin{itemize}
\item The method that simply assumes the series has the wanted convergent behavior and then directly uses the information of latest known $n_{\rm th}$-order term as an estimation of the contribution of $(n+1)_{\rm th}$-order term, e.g, the $(n+1)_{\rm th}$-order term is assumed to be $\pm |C_{n}(\mu_r)\alpha_s^{n+1+l}(\mu_r)|_{\rm MAX}$ or more conservatively, $\pm |C_{n}(\mu_r)\alpha_s^{n+l}(\mu_r)|_{\rm MAX}$, where ``MAX'' denotes the maximum value obtained by varying $\mu_r$ within the chosen range;

\item The resummation method that involves introducing an appropriate generating function, which can generate the same given series. The generating function builds a bridge between the series and the functions using the formal $\alpha_s$-power series, which allows us to use functional analysis to investigate the limiting properties of series~\cite{Herbert:1990}. The frequently used generating function is the fractional generating function and its variants, which is commonly referred to as the Pad$\acute{e}$ approximation approach~\cite{Basdevant:1972fe, Samuel:1992qg, Samuel:1995jc, Brodsky:1997vq, Boito:2018rwt}. There are some other resummation methods, such as the one by conformal mapping of Borel plane~\cite{Caprini:2019kwp} and the one by contour-improved perturbation theory~\cite{Beneke:2008ad, Beneke:2012vb}. Unlike the generating function method, they re-summed $\alpha_s$ into another variable to obtain its limiting behavior, and finally restored it to the expansion of $\alpha_s$ to obtain the UHO-terms estimations;

\item The probability method that suggests to quantify the UHO-contribution in terms of a proper probability distribution, such as the one constructed via the Bayesian analysis~\cite{Duhr:2021mfd, Cacciari:2011ze, Bonvini:2020xeo}. Practically, such trend can be considered as a feedback from the known series, and the precision of the probability distribution can be enhanced iteratively.
\end{itemize}

It has been observed that the feasibility, reliability, and precision of any method used to estimate the UHO contribution from a given fixed-order pQCD series are highly contingent upon the precision and convergent behavior of the series. And to improve the predictive power of any method for achieving the UHO contributions, or equivalently to improve the precision of the predicted UHO contributions, it is important to have a more convergent and more precise pQCD series as much as possible. Practically, one may achieve more precise estimation of the UHO contributions when more loop terms become available. However due to the complexity of loop calculations, it is important to find a more reliable method that can work effectively with less known loop terms.

Mathematically, the regression analysis is a quantitative method used to examine the relationship between the variables of interests and the other variables. Depending on the chosen model for generating function, various types of regression analyses have been performed in the literature, such as the linear regression, the logistic regression, and etc. Among these, the linear regression method is an important and useful tool frequently employed in applied statistics~\cite{Hocking:1996, Eisenhauer:2003, Olive:2017}. Practically, one may first introduce an initial generating function to simulate the asymptotic behavior of a given fixed-order series, which will then be tamed via a step-by-step way and become more accurate when more loop terms have been applied to update its input parameters. In previous literature, the linear regression has been employed for predicting the black hole masses using reverberating active galactic nucleus samples~\cite{Chainakun:2022lma} and for estimating the Higgs couplings within the standard model effective field theory~\cite{Murphy:2017omb}. In this manuscript, we will propose a novel method for estimating the UHO contributions of the fixed-order pQCD series.  The method is grounded in the convergence properties of asymptotic series prior to optimal truncation. Specifically, we model the dominant exponential behavior of the series coefficients, which leads to a linear relation upon taking the logarithm. This allows the estimation problem to be reduced to one of linear regression for the characteristic slope parameter. Under appropriate modifications, we refer to this approach as linear regression through the origin (LRTO).

A physical observable should satisfy renormalization group invariance (RGI)~\cite{Gell-Mann:1954yli, Callan:1970yg, Symanzik:1970rt, StueckelbergdeBreidenbach:1952pwl, Peterman:1978tb, Wu:2014iba}. However satisfying the RGI is a challenging problem for a fixed-order pQCD approximant, since the truncated series (\ref{generalpQCD}) does not automatically satisfy the RGI. Especially, a pQCD series based on a guessed choice of renormalization scale is scheme-and-scale dependent, leading to conventional scheme-and-scale ambiguities. This is because a simple assignment of the renormalization scale results in a mismatch between the QCD running coupling and the corresponding perturbative coefficients at each order. Furthermore, we are also uncertain about the range over which scale and scheme parameters should vary to achieve a reasonable error estimation. A more convergent pQCD series with more loop terms are beneficial for softening conventional scheme-and-scale ambiguities~\cite{Wu:2013ei, Wu:2019mky, DiGiustino:2023jiq}. However even if the net value (the total cross-section or the total decay width) of the series becomes smaller with increasing loop terms, which is due to large cancellation of scale dependence among different orders, the magnitude of each order terms itself is still highly uncertain; Thus, the UHO contribution derived from this initial scale-dependent series will be questionable. For later convenience, we will also refer to this {\bf initial} series under the above mentioned conventional scale-setting treatment as the {\bf conventional} series. In order to achieve a precise UHO contribution, it is crucial to find a proper scale-setting approach to improve the initial pQCD series so as to establish a more reliable and more precise fixed-order pQCD prediction that is free of any choices of renormalization scale and scheme.

In different to other scale-setting approaches suggested in the literature, the Principle of Maximum Conformality (PMC) offers a rigorous solution for conventional scheme-and-scale ambiguities~\cite{Brodsky:2011ta, Brodsky:2012rj, Brodsky:2011ig, Mojaza:2012mf, Brodsky:2013vpa}, which extends the well-known Brodsky-Lepage-Mackenzie (BLM) method~\cite{Brodsky:1982gc} from one-loop level up to all orders by systematically utilizing the renormalization group equation (RGE) for fixing the $\alpha_s$-running behavior of the considered process. In $N_C\to 0$ Abelian limit~\cite{Brodsky:1997jk}, the PMC procedure is equivalent to the Gell-Mann-Low (GM-L) procedure used for quantum electrodynamics~\cite{GellMann:1954fq}. It has been demonstrated that the PMC can eliminate the conventional scheme-and-scale ambiguities simultaneously, being in good accordance with the self-consistency requirements of renormalization group~\cite{Brodsky:2012ms} and the standard RGI~\cite{Wu:2018cmb}. Furthermore, the commensurate scale relations among different pQCD approximants also ensure the scheme independence of the PMC predictions~\cite{Brodsky:1994eh, Huang:2020gic}. In accordance with the PMC procedures, the RGE-involved non-conformal $\{\beta_i\}$-terms will be eliminated from the pQCD series and be used to fix the correct/effective magnitude of $\alpha_s$ of the process, and the divergent renormalon terms~\cite{Beneke:1994qe, Neubert:1994vb, Beneke:1998ui} that are proportional to the $n! \beta_0^n \alpha_s^n$-like terms (using the large $\beta_0$-approximation $\beta_i \sim \beta_0^{i+1}$) are eliminated accordingly, naturally leading to a more convergent perturbative series. Such improvement of pQCD convergence has been confirmed by numerous PMC applications done in the literature~\footnote{In a few instances, the conformal coefficients are substantially large, and the $\alpha_s$ power suppression cannot catch up with the increased magnitude of the conformal coefficient at higher-orders. As a result, the series may not exhibit the desired convergent behavior even after applying the PMC. This divergent behavior can be considered as one of the intrinsic properties of the pQCD series, which does not affect the applicability of PMC.}. 

It has been demonstrated that the PMC scheme- and scale-invariant series is valuable for enhancing the accuracy of the Pad$\acute{e}$ approximation approach~\cite{Du:2018dma} and the Bayesian approach~\cite{Shen:2022nyr} in estimating UHO contributions. In this paper, we utilize both the initial pQCD series and the improved pQCD series -- derived via the PMC single-scale approach (PMCs)~\cite{Shen:2017pdu, Yan:2022foz} -- to test the effectiveness of the presently proposed LRTO approach. We present a comparison of the LRTO results obtained for the scale-dependent initial series and the PMC series. As shown, the scale-invariant and more convergent PMC series is indeed better suited to achieving the precision goal of estimating UHO contributions: it exhibits superior convergence properties that enable faster and more stable predictions. Moreover, it effectively reduces the residual scale dependence associated with UHO terms~\footnote{While the PMC series is scale- and scheme-invariant, both it and the PMC scale still exhibit perturbative characteristics that result in residual scale dependence due to UHO terms~\cite{Zheng:2013uja}.}~\footnote{Additionally, there are \(\{\beta_i\}\)-terms arising from the renormalization of the heavy quark's mass and wave function, among others, which are unrelated to \(\alpha_s\) renormalization. These should be preserved as conformal coefficients to fix the correct magnitude of \(\alpha_s\), though this can be addressed separately~\cite{Huang:2022rij, Ma:2024xeq, Yan:2024oyb}.}.

The remaining parts of the paper are organized as follows. In Section~\ref{secII}, for self-consistency, we first provide a brief introduction to the linear regression method, following the standard representation from the Particle Data Group~\cite{Workman:2022ynf}. In Section~\ref{secIII}, we demonstrate how to apply this method to a fixed-order pQCD series to obtain its UHO contribution. In Section \ref{secIV}, we present numerical results for important ratio $R_\tau$, obtained using the LRTO method. The respective pQCD series are treated under both the conventional scale-setting approach and the PMCs scale-setting approach. Section~\ref{secV} is reserved for a summary.
	
\section{Basic definitions of the linear regression method} \label{secII}

Consider a set of $n$ measurements ${y_i}$ taken at known points ${x_i}$, where $i\in[1,n]$. Each measurement $y_i$ is assumed to follow a Gaussian distribution with a mean of $\mu(x_i,\bm{\theta })$ and a known variance of $\sigma_i$. Here, the symbol $\bm{\theta }$ represents the parameter vector, where its dimension $m$ indicates the number of parameters employed to characterize the relationship between $x_i$ and $y_i$. An estimator for an arbitrary parameter $\hat{\phi}$ (denoted with a hat) is defined as a function of the data employed to estimate the value of the parameter $\phi$. The objective of this method is to construct the estimators $\hat{\bm{\theta}}$ for the unknown parameters within $\bm{\theta }$. Statistically, an important way to estimate the values of the parameters $\bm{\theta}$ is the method of least squares (LS), which determines the parameters by minimizing $\chi^2$. Since, in the present case, the LS method coincides with the method of maximum likelihood, and the LS estimator can be determined by minimizing the log-likelihood function with the covariance matrix $V_{ij}=\text{cov}[y_i,y_j]$, which can be expressed as
\begin{align}
	\chi^2(\bm{\theta})=\big(\bm{y}-\bm{\mu}(\bm{\theta})\big)^T V^{-1}\big(\bm{y}-\bm{\mu}(\bm{\theta})\big),
\end{align}
where $\bm{y}=(y_1,\cdots,y_n)$ is the column vector of measurements, $\text{cov}[y_i,y_j]$ represents the covariance between $y_i$ and $y_j$, $\bm{\mu(\theta)}$ is the corresponding vector of predicted values, and the superscript $T$ denotes the transpose.

In the linear regression method, $\mu(x_i,\bm{\theta})$ is a linear function of the parameters, i.e.,
\begin{align}
	\mu(x_i,\bm{\theta})=\sum_{j=1}^{m}\theta_j h_j(x_i),
\end{align}
where $m\leq n$, and $h_j(x_i)$ are $m$ linearly independent functions. Defining $H_{ij}=h_j(x_i)$ and setting derivatives with respect to the $\theta_i$ of $\chi^2$ equal to zero can minimize $\chi^2$ and gives the LS estimators,
\begin{align}
	\hat{\bm{\theta}}=(H^T V^{-1} H)^{-1} H^T V^{-1}\bm{y},  \label{thetabm}
\end{align}
where the estimator $\hat{\bm{\theta}}$ (written with a hat) is a function of the data used to estimate the values of the parameters within $\bm{\theta}$. The covariance matrix of $\bm{\theta}$ for the estimator, $U_{ij}=\text{cov}[\hat{\theta_i},\hat{\theta_j}]$, can be written as
\begin{align}
	U=(H^T V^{-1} H)^{-1}.
	\label{lRV}
\end{align}

In the linear regression, the variance of $y_i$ is equal, and each $y_i$ is independent, which means the elements of the matrix $V$ satisfy
\begin{align}
	V_{ij}=\sigma^2 \delta_{ij}, 
\end{align}
where $\delta_{ij}=1$ for $i=j$, $\delta_{ij}=0$ for $i\neq j$. Eq. (\ref{thetabm}) and Eq. (\ref{lRV}) can be rewritten as
\begin{equation}
	\hat{\bm{\theta}}=(H^T H)^{-1} H^T\bm{y}, \quad U=\sigma^2 (H^T H)^{-1}.
\end{equation}
It shows that the estimator $\hat{\bm{\theta}}$ does not depend on $\sigma$. Thus, if $\sigma^2$ is not known, the parameters can be estimated by using the method of LS in the linear regression. After acquiring the estimator $\hat{\bm{\theta }}$, the unbiased estimation of $\sigma^2$ can be expressed in term of residuals between measurements $y_i$ and the regression function $\mu(x_i,\bm{\hat{\theta }})$ as~\cite{Olive:2017}
	\begin{align}
		\hat{\sigma}^2=\frac{1}{n-m}\big(\bm{y}-\bm{\mu}(\bm{\hat{\theta }})\big)^T\big(\bm{y}-\bm{\mu}(\bm{\hat{\theta }})\big).
	\end{align}

\section{Linear Regression in pQCD and the magnitude of UHO terms} \label{secIII}

In this section, we demonstrate how the concept of optimal truncation for asymptotic series in perturbative QCD, coupled with a linear regression analysis, can be employed to estimate the behavior of uncalculated higher-order (UHO) terms and quantify their potential contribution to a physical observable.
 
The basic element of our estimation method is the mathematical nature of the perturbation series as an asymptotic series. For such a series, there exists an optimal truncation order $N^{*}$ at which the perturbative approximation to the true value $\rho$ is most accurate. The truncation error for the series with highest correction order $N < N^{*}$ is then bounded by:
\begin{equation}
	\left| \rho - \sum_{k=0}^{N} C_k \alpha_s^{k+l} \right| < L_{N} \, \alpha_s^{N+1+l},
\end{equation}
where $L_N$ is a positive constant. This inequality defines the range $N < N^{*}$ within which the series maintains convergence, primarily governed by the exponential suppression from powers of the coupling constant $\alpha_s$. Because the divergent behavior of the expansion coefficients is mainly caused by the so-called renormalon terms, it has been roughly suggested that $N^{*} \sim 1/\alpha_s$~\cite{Beneke:1998ui}, where the factorial growth in the renormalon-terms overcomes the power suppression of $\alpha_s$. Because the RGE-involved $\{\beta_i\}$-terms have been adopted to fixing the effect magnitudes of the scale-running parameters such as $\alpha_s$, the PMC can removes such kind of renormalon-like terms by recursively using of RGE, thus it not only leads to a better convergent series, but also effectively delays the onset of the optimal truncation order $N^*$, thereby extending the predictive range of the fixed-order expansion within its inherent asymptotic nature~\footnote{The PMC procedure is a type of $\{\beta_i\}$-resummation method designed to obtain an accurate fixed-order perturbative series. It exclusively eliminates renormalon terms associated with the renormalization group equation and is independent of all other types of divergent terms. As a result, it can improve the convergence of the series to a certain extent, yet it cannot alter the intrinsic nature of the series.}.

It is advantageous to define a normalized $K$-factor that isolates the relative size of successive perturbation contributions:
\begin{align}
	K_k = \frac{C_k}{C_0} \alpha_s^k, \quad (k \geq 1).
	\label{kfactor}
\end{align}
Using this $K$-factor, we rewrite the given $n_{\rm th}$-order series \eqref{generalpQCD} as
\begin{equation}
	\rho_n=C_0\alpha_s^l \left(1+\sum_{k=1}^{n-1} K_k  \right).
\end{equation}

It is noticed that the magnitude of the $K$-factor, prior to the optimal truncation order $N^*$, is governed by $\alpha_s$ exponential suppression. Therefore we extract the exponential behavior as a parameter $q$ to characterizing the convergence rate. And the minor deviations (sub-leading terms) from this pure exponential trend are absorbed into a factor $\epsilon_k$.
\begin{equation}
	\left| K_k \right| = q^{\,k} \exp(\epsilon_{k}).
	\label{kmodel}
\end{equation}
Taking the natural logarithm yields a linear relation:
\begin{equation}
	\ln |K_k| = k \cdot \theta + \epsilon_{k},
	\label{linearRelation}
\end{equation}
where $\theta = \ln q < 0$. The slope parameter $\theta$ thus encapsulates the primary convergence rate of the perturbation series before the onset of asymptotic divergence. Consequently, the task of estimating the trend of the unknown higher-order $|K_k|$ is reduced to that of estimating the magnitude of the linear parameter $\theta$.

Furthermore, to apply linear regression, we assume the minor deviation terms $\{\epsilon_k\}$ for the known low-order terms ($k=1, ..., n$) are independent and identically distributed random variables following a normal distribution with zero mean and constant variance $\sigma^2$:
\begin{equation}
 f(\epsilon_k;0,\sigma)=\frac{1}{\sqrt{2\pi}\sigma} \exp\left(-\frac{\epsilon_k^2}{2\sigma^2}\right).
\end{equation}

This assumption allows us to treat the deviations of the known $\ln |K_k|$ from a perfect line as normally distributed noise. With these two assumptions, the known low-order K-factor $\{K_1, ..., K_n\}$ provide the known points and measurements $(k, \ln|K_k|)$. Performing a linear regression on this data yields the best-fit estimate for the slope $\hat{\theta}$ and its standard error $\delta$. The fitted model, $\ln |K_k| \approx k \cdot \hat{\theta}$, can then be extrapolated to estimate the magnitude of the next unknown term ($K_{n+1}$) , thereby providing a quantitative estimate for the remaining theoretical uncertainty.

We substitute the order $k$ as the known point $x_k$ in the linear regression and the logarithm of the $K$-factor $\ln|K_k |$ as the measured quantity $y_k$ in the derivation of Sec. \ref{secII}. Then we can derive the estimators for $\theta$ and $\sigma$ as follows,
\begin{align}
	\hat{\theta} &=\frac{6}{(2n-1)(n-1)n}\sum_{k=1}^{n-1}k\ln|K_k|,
	\label{e2}\\
	\hat{\sigma}^2&=\frac{1}{(n-2)}\sum_{k=1}^{n-1}\left(\ln|K_k|- k\cdot\hat{\theta} \right)^2.
	\label{e3}
\end{align}
Further, the variance $\delta$ of $\theta$ can be derived with the help of Eq.~(\ref{lRV}):
\begin{align}
	\delta^2=\frac{6}{(2n-1)(n-1)n}\hat{\sigma}^2.
	\label{delta}
\end{align}
This particular regression without intercept distance is the so-called linear regression through the origin (LRTO). Then logically, the probability density function (p.d.f.) of $\theta$ can be derived via the following way:
\begin{align}
	f(\theta ;\hat{\theta},\delta)=\frac{1}{\sqrt{2\pi}\delta} \exp\left[-\frac{(\theta -\hat{\theta})^2}{2\delta^2}\right],
	\label{theta}
\end{align}
where the parameter $\hat{\theta}$  represents the rate of convergence. The parameter $\delta$ reacts the uncertainty of $\theta$ which is introduced by the higher-order minors in Assumption 1.

So far, the credible interval (CI) of $\theta$ can be confirmed at a fixed degree-of-belief (DoB) $p\%$. For a standard Gaussian distribution, the CI about a DoB is defined as
    \begin{align}
    	p\%=\int_{-u_{p/2}}^{u_{p/2}}\frac{1}{\sqrt{2\pi}}\exp\left(-\frac{x^2}{2}\right) \mathrm{d}x.
    	\label{sND}
    \end{align}
When $p=68.3$, $95.5$ and $99.7$, the $u_{p/2}$ will equal to $1$, $2$ and $3$, respectively. The CI of $\theta$ can be given as
    \begin{align}
    	\theta ^{(p)}&\in[\hat{\theta}-u_{p/2}\delta,\hat{\theta}+u_{p/2}\delta],
    	\label{CItheta}\\
    \end{align}
by transformation $x\to(x-\hat{\theta})/\delta$ in \eqref{sND}. And the CI of the convergence rate can be given as 
\begin{align}
	  q^{(p)}\in[\hat{q}\exp(-u_{p/2}\delta),\hat{q}\exp(u_{p/2}\delta)],
\end{align}
 where $\hat{q}=\exp(\hat{\theta})$.

Using Eqs.(\ref{generalpQCD}, \ref{kfactor}, \ref{kmodel}), one can derive the p.d.f. for the $K$-factors $K_i$, the expansion coefficients $C_i$ and $\rho_n$, respectively. Their p.d.f obey the Log-Gaussian distribution:
    \begin{align}
    	L_G(x;\mu,\sigma)=\frac{1}{\sqrt{2\pi}\sigma}\frac{1}{x}\exp\left(-\frac{(\ln x-\mu)^2}{2\sigma^2}\right), \,\, x>0.
    	\label{LGD}
    \end{align}
Then the p.d.f. of $K_i$, $C_i$ and $\rho_{n+1}$ are
    \begin{align}
    	f(|K_k|)&=L_G(|K_k|;\hat{\theta}k,\delta k),  \label{fki}\\
    	f(|C_k|)&=L_G\left(|C_k|;(\hat{\theta}-\ln\as)k+\ln|C_0|,\delta k\right),  \label{fci}\\
    	f(\rho_{n+1})&=\frac{1}{2}L_G\left(|\rho_{n+1}-\rho_n|;\hat{\theta}n+\ln|C_0\as^{l}|,\delta n\right).   \label{frho}
    \end{align}
From the correspondence between the Gaussian distribution and the log Gaussian distribution, we then obtain the CIs of $K_k$ and $C_k$ by using Eqs.(\ref{kmodel}, \ref{CItheta}):
    \begin{align}
    	|K_{k}|^{(p)}&\in \left[\exp{(\hat{\theta}k-u_{p/2}k\delta)},\exp{(\hat{\theta}k+u_{p/2}k\delta)}\right],
    	\label{CIK}\\
    	|C_{k}|^{(p)}&\in \left[|C_0|\as^{-k}e^{(\hat{\theta}k-u_{p/2}k\delta)},|C_0|\as^{-k}e^{(\hat{\theta}k+u_{p/2}k\delta)}\right].
    	\label{CIC}
    \end{align}

To acquire the CI of $\rho_{n+1}$, the $|K_k|^{(p)}$ needs to be shifted to $K_k^{(p)}$ as the estimation of UHO-terms. Considering the worst condition, the upper of $|K_k|^{(p)}$ is taken as the boundary of $K_k^{(p)}$, marked as $\Delta_{n}^{(p)}$:
    \begin{align}
    	\Delta_{n}^{(p)}=|C_0|\as^l\exp{(\hat{\theta}n+u_{p/2}n\delta)},
    	\label{uho}
    \end{align}
and then the CI of $\rho_{n+1}$ can be obtained via the following way,
    \begin{align}
    	\rho_{n+1}^{(p)}\in[\rho_{n}-\Delta_n^{(p)},\rho_{n}+\Delta_{n}^{(p)}].
    	\label{CIrho}
    \end{align}
Using Eq.\eqref{sND}, it is found that the actual DoB of $\rho_{n+1}$ is $(p/2+50)\%$. To be consistent with the naming of the other perturbation parameters, for simplicity and without introducing any confusion, we will still refer to the DoB of this CI as $p\%$.
	
As a final remark, since the LRTO method is a kind of linear regression method, one may need to judge whether the LRTO method is applicable to a pQCD series. For a pQCD series known up to $n_{\rm th}$-order QCD corrections, we can introduce the following determination coefficient ${\cal R}$ for the purpose, e.g.~\cite{Olive:2017, Eisenhauer:2003}
	\begin{equation}
		{\cal R}^2=\frac{\sum\limits_{k=1}^{n-1}k^2\hat{\theta}^2}{\sum\limits_{k=1}^{n-1}\ln^2|K_k|}.
	\end{equation}
The coefficient of determination, ${\cal R}$, is used to evaluate the goodness-of-fit. For example, if we require that there is a $95.5\%$ certainty that this estimation is reliable, then we must have ${\cal R}>0.997$, $0.950$ and $0.878$ for the case of $n=2$, $3$ and $4$, respectively. For definiteness, in this paper, the significance level of $95.5\%$ is selected to do our calculation and discussions.

\section{Explicit example using the LRTO method, $R_\tau$} \label{secIV}

\subsection{A brief introduction of PMC single-scale setting approach}

Before proceeding, we first provide a brief introduction to the PMCs approach; detailed formulas for this method can be found in Refs.~\cite{Shen:2017pdu, Yan:2022foz}. The idea of the single-scale setting procedure has been initially suggested in Refs.\cite{Grunberg:1991ac, Brodsky:1995tb}, which attempts to extend the BLM method up to two-loop QCD corrections by using the $n_f$-series as the starting point. Lately, it has been found that by transforming the $n_f$-series into $\{\beta_i\}$-series correctly with the help of RGE, a self-consistency extension of BLM up to all-orders can be achieved~\cite{Shen:2017pdu}. It has been shown that the PMCs approach can be served as a reliable substitute for the strict multi-scale PMC approach~\cite{Brodsky:2011ta, Brodsky:2012rj, Brodsky:2011ig, Mojaza:2012mf, Brodsky:2013vpa}. And it does lead to more precise pQCD predictions with less residual scale dependence, cf. a recent comparison of various PMC scale-setting procedures in Ref.\cite{Huang:2021hzr}. The PMCs also provides a self-consistent way to achieve precise $\alpha_s$-running behavior in both the perturbative and nonperturbative domains~\cite{Deur:2017cvd, Yu:2021yvw}. The PMCs approach adopts the RGE to fix an overall effective magnitude of $\alpha_s$ and uses the QCD degeneracy relations~\footnote{It has been demonstrated that the degeneracy relations are general property of QCD theory~\cite{Bi:2015wea}, which also confirms~\cite{Shen:2016dnq} the correctness of the well-known generalized Crewther relation~\cite{Broadhurst:1993ru, Crewther:1997ux}.} among different orders to transform the RGE-involved $n_f$-series of the initial pQCD series into the $\{\beta_i\}$-series. And following standard PMCs procedures, the pQCD approximant $\rho_n$ known up to $n_{\rm th}$-order QCD corrections, e.g. Eq.(\ref{generalpQCD}), can be reformulated in the following manner
\begin{widetext}
\begin{align}		
\rho_{n}=&\,r_{1,0}\alpha_{s}^{l}(\mu_{r})+\left[r_{2,0}+l\beta_{0}r_{2,1}\right]\alpha_{s}^{l+1}(\mu_{r})  +\bigg[r_{3,0}+l\beta_{1}r_{2,1}+(l+1)\beta_{0}r_{3,1} +\frac{l(l+1)}{2}\beta_{0}^{2}r_{3,2}\bigg]\alpha_{s}^{l+2}(\mu_{r})\notag\\	&+\left[r_{4,0}+l\beta_{2}r_{2,1}+(l+1)\beta_{1}r_{3,1}+\frac{l(3+2l)}{2}\beta_{0}\beta_{1}r_{3,2} +(l+2)\beta_{0}r_{4,1} \right. \notag\\
&\left. \quad +\frac{(l+1)(l+2)}{2}\beta_{0}^{2}r_{4,2}+\frac{n(n+1)(n+2)}{3!}\beta_{0}^{3}r_{4,3}\right] \alpha_{s}^{l+3}(\mu_{r})\notag\\
& +\left[r_{5,0}+l\beta_{3}r_{2,1}+(l+1)\beta_{2}r_{3,1}+\frac{l(l+2)}{2}\left(\beta_{1}^{2} +2\beta_{0}\beta_{2}\right)r_{3,2}+(l+2)\beta_{1}r_{4,1} \right.\notag\\
& \left. \quad +\frac{(l+1)(2l+5)}{2}\beta_{0}\beta_{1}r_{4,2} +\frac{l(3l^{2}+12l+11)}{6}\beta_{0}^{2}\beta_{1}r_{4,3} +(l+3)\beta_{0}r_{5,1} \right.\notag\\
& \left. \quad +\frac{(l+2)(l+3)}{2}\beta_{0}^{2}r_{5,2} +\frac{(l+1)(l+2)(l+3)}{6}\beta_{0}^{3}r_{5,3}  +\frac{l(l+1)(l+2)(l+3)}{24}\beta_{0}^{4}r_{5,4}\right]\alpha_{s}^{l+4}(\mu_{r}) +\cdots,
\label{pQCD approximant}
\end{align}
\end{widetext}
where for simplicity, we do not write down the explicit $\mu_r$-dependence in the expansion coefficients. And the $\mu_r$-dependence coefficients $r_{i,j}$ can be redefined as
\begin{equation}\label{rij-hatrij}
	r_{i,j}=\sum_{k=0}^{j}C_{j}^{k}\hat{r}_{i-k,j-k}\ln^{k}\frac{\mu_{r}^{2}}{Q^{2}},
\end{equation}
where $C_{j}^{k}=j!/\left(k!(j-k)!\right)$ are combination coefficients, and $\hat{r}_{i,j}=r_{i,j}|_{\mu_{r}=Q}$. Specially, we have $\hat{r}_{i,0}\equiv r_{i,0}$. Following the standard PMCs procedures, an overall effective coupling -- and hence an overall effective scale -- can be determined by requiring all non-conformal $\{\beta_i\}$-terms at each order to vanish. And then, the pQCD approximant (\ref{generalpQCD}) changes to the following conformal series
\begin{align}
	\rho_n\big|_{\text{PMCs}}=\sum_{k=0}^{n-1}\hat{r}_{k+1,0}\as^{l+k}\left(Q_*\right),
	\label{PMCs}
\end{align}
where $\hat{r}_{k+1,0}$ are conformal coefficients. For a pQCD series known up to $n_{\rm th}$-order QCD corrections, the PMC effective scale $Q_*$ can be determined up to next-to-$\cdots$-next-to-leading log accuracy ($\mathrm{N^{n-2}LL}$-accuracy); or more explicitly, the logarithmic term $\ln{Q_*^2/Q^2}$ can be expanded as a power series in terms of $\as(Q_*)$:
\begin{align}
	\ln \frac{Q_*^2}{Q^2}=\sum_{k=0}^{n-3}S_k\as^k(Q_*),
	\label{Qs}
\end{align}
where the coefficients $S_i$ are functions of the given coefficients $C_i(Q)$. The expressions of those coefficients, together with the conformal coefficients $\hat{r}_{i,0}$, can be found in Ref.\cite{Yan:2022foz}. The right-hand side of Eq.(\ref{Qs}) is also a power series in $\as$, indicating the perturbative nature of the PMC scale ($Q^{*}$). It is a kind of resummation, which resums all the known type of $\{\beta_i\}$-terms of the pQCD series (\ref{pQCD approximant}) and determines the precise magnitude of $\alpha_s$. The determined scale $Q_*$ is generally different from the usually chosen typical momentum flow $Q$ of the process. For example, from the above relation between $Q^{*}$ and $Q$, one can easily obtain the well-known one-loop relation for the strong coupling constant~\cite{Brodsky:1982gc}, $\alpha_s^{\overline{\rm MS}}(e^{-5/6}Q) =\alpha_s^{\rm GM-L}(Q)$, where the scale displacement parameter $e^{-5/6}$ between the $\overline{\rm MS}$ scheme and the GM-L scheme~\cite{GellMann:1954fq} is a result of the convention that is chosen to define the minimal dimensional regularization scheme~\cite{Bardeen:1978yd}. Hence the PMC scale $Q_*$ can be treated as the (correct) effective momentum flow of the process. Together with the $\mu_r$-independent conformal coefficients, the resulting pQCD series is exactly scheme and scale independent, thus providing a reliable basis for estimating the contributions of the uncalculated UHO-terms.

\subsection{Numerical results and discussions}

In this subsection, we will apply the LRTO method to deal with pQCD approximant of the physical observable $R_\tau$, showing how the LRTO method works.

The ratio $\Rti(M_\tau)$ is defined as 
\begin{align}
\Rti(M_\tau) =&\frac{\sigma(\tau\to\nu_\tau+\text{hadrons})}{\sigma(\tau\to\nu_\tau+\bar{\nu}_ e+e^-)} \notag\\
=&3\sum|V_{ff'}|^2 \left(1+\tilde{R}_n(M_\tau)\right),
\end{align}
where $V_{ff'}$ are Cabbibo-Kobayashi-Maskawa matrix elements, $\sum\left|V_{ff'}\right|^2 =\left|V_{ud} \right|^2+\left|V_{us}\right|^2\approx 1$ and $M_{\tau}= 1.77686$ GeV \cite{Workman:2022ynf}. The QCD corrections of $\Rti(M_\tau)$, denoted by $\tilde{R}_{n}(M_{\tau})$, can be written as
\begin{align}
	\tilde{R}_n(M_\tau) &=\sum_{k=1}^{n} C_k(\mu_r/M_\tau)\alpha_s^k(\mu_r).
\end{align}
Here the perturbative coefficients $C_k$ for the initial pQCD series and the corresponding coefficients $r_{i,j}$ for the resultant PMCs series can be obtained by using the known relation of $\Rti(M_\tau)$ to $R_{e^+e^-}(Q)$~\cite{Lam:1977cu}.

In performing the numerical analyses, the running of $\alpha_s$ has been assumed at the four-loop level, and the RunDec program~\cite{Chetyrkin:2000yt, Herren:2017osy} is adopted. And the QCD asymptotic scale $\Lambda_{\text{QCD}}$ is fixed by using $\alpha_s(M_Z)=0.1179$~\cite{Workman:2022ynf}, giving $\Lambda_{\text{QCD}}^{\{n_f=5\}}=0.210$ GeV.  

First, we show how the PMC scale $Q_*$ changes as more loop corrections are included. Since the RGE-involved $\{\beta_i\}$-terms of $\tilde{R}_n(M_\tau)$ starts at the two-loop level, by applying the standard PMCs procedures, we can fix the PMC scale $Q_*$ up to LL-, NLL- and NNLL accuracy by using the two-loop, three-loop and four-loop QCD corrections, respectively. Because the perturbative series of $\ln Q_{*}^{2}/M^{2}_{\tau}$ presents good convergent behavior, the PMC scale $Q_*$ converges as more loop corrections are included. More explicitly, we have 
\begin{align}
	Q^{\rm LL}_* &= 0.90 \; {\rm GeV}, \\
	Q^{\rm NLL}_* &= 1.01 \; {\rm GeV}, \\
	Q^{\rm NNLL}_* &= 1.07 \; {\rm GeV}. 
\end{align}
The results show a monotonic increase: $Q^{\rm LL}_* < Q^{\rm NLL}_* < Q^{\rm NNLL}_*$, and the difference between the two nearby values becomes smaller and smaller when more loop-terms are included, e.g. $|Q^{\rm NLL}_{*} - Q^{\rm LL}_{*}|<|Q^{\rm NNLL}_{*} - Q^{\rm NLL}_{*}|$.

\begin{table}[htb]
	\centering
	\begin{tabular}{c|cccc}
		\hline
		~~	&$k=1$&$k=2$&$k=3$&$k=4$\\
		\hline
		$C_k\,(\mu_r=M_\tau)$ & $0.3183$ & $0.5271$ & $0.8503$ & $1.3046$ \\
		$C_k\,(\mu_r=2 M_\tau)$ & $0.3183$ & $0.7968$ & $2.117$ & $5.7141$ \\
		$C_k\,(\mu_r=4 M_\tau)$ & $0.3183$ & $1.0197$ & $3.4589$ & $11.9181$ \\
		$\hat{r}_{k,0}$ & $0.3183$ & $0.2174$ & $0.1108$ & $0.0698$ \\
		\hline
	\end{tabular}
	\caption{The known expansion coefficients of $\tilde{R}_{n=4}(M_\tau)$ before and after applying the PMCs. The scale-dependent coefficients $C_k(\mu_r)$ of the initial series are for $\mu_r=M_\tau$, $\mu_r=2 M_\tau$, and $\mu_r=4 M_\tau$, respectively. The PMCs conformal coefficients $\hat{r}_{k,0}$ are scale-independent. }
	\label{taucoefficient}
\end{table}

Second, we show how the perturbative coefficients change before and after applying the PMCs scale-setting approach. We present the calculated coefficients for the four-loop result of $\tilde{R}_{4}(M_\tau)$ in Table  \ref{taucoefficient}, e.g. the first four conformal coefficients $\hat{r}_{k,0}$ with $k=(1,\cdots,4)$ for the PMCs series and the first four coefficients $C_k$ for the initial pQCD series. Table  \ref{taucoefficient} shows the coefficients $C_k$ of the initial pQCD series are highly scale dependent. It shows that the coefficients of the initial series increases for the higher-order terms, which will become larger as the scale is more different from the usual choice of $\mu_r=M_\tau$. Following this arising trends, one may expect that at enough higher orders, their coefficients may rightly cannel the $\alpha_s$ power suppression, leading to a divergent behavior; This divergent behavior is usually called by the renormalon terms in the expansion coefficients~\cite{Beneke:1994qe, Neubert:1994vb, Beneke:1998ui}. This fact also indicates that the usual ``guessed'' choice of $\mu_r=M_\tau$ to eliminate the large log terms proportional to certain powers of $\ln\mu_r^2/M_\tau^2$ can only partly remove the divergent terms. Moreover, it is found that the expansion coefficients at every orders are highly scale dependent for the initial series. Thus the LRTO approach can only be applied after one specifies the choice of $\mu_r$, and different choices of $\mu_r$ will introduce extra uncertainty for any approach of estimating the contribution of UHO-terms. For definiteness, in the following, we will adopt $\mu_r \equiv M_\tau$ to do the LRTO analysis for the initial scale-dependent fixed-order series.

On the other hand, the PMCs series does not have such renormalon divergence and the conformal coefficients $\hat{r}_{k,0}$ shown in Table \ref{taucoefficient} decrease with the increment of the loop terms, leading to a much more convergent series. Since the divergent $\{\beta_i\}$-terms of the initial series have been used to fix the correct magnitude of the strong coupling constant, the resulting conformal PMCs series is not only scale-independent but also more convergent. Thus the PMCs series will be a more reliable and more precise basis for estimating the UHO contributions.

\begin{table}[htb]
	\centering
	\begin{tabular}{c|cccc}
		\hline
		order-$n$ & $\hat{\theta}$ & ${\cal R}$ & $\delta$ & $\hat{q}$ \\
		\hline
		2&$-1.533$&$0.9951$&$0.1527$&$0.2156$\\
		3&$-1.566$&$0.9978$&$0.0738$&$0.2090$\\
		4&$-1.511$&$0.9983$&$0.0512$&$0.2221$\\
		\hline
	\end{tabular}
	\caption{The estimator $\hat{\theta}$, the determination coefficient ${\cal R}$, the variance $\delta$ and the convergence rate $\hat{q}$ for LRTO approach by using the PMCs series of $\tilde{R}'_n(M_\tau)$ up to  $n_\text{th}$-order QCD corrections, where $n=(2,3,4)$, respectively. }
	\label{t3}
\end{table}

\begin{table}[htb]
	\centering
	\begin{tabular}{c|cccc}
		\hline
		order-$n$ & $\hat{\theta}$ & ${\cal R}$ & $\delta$ & $\hat{q}$ \\
		\hline
		2&$-1.642$&$0.9804$&$0.3298$&$0.1936$\\
		3&$-1.353$&$0.9785$&$0.2091$&$0.2584$\\
		4&$-1.204$&$0.9810$&$0.1376$&$0.3000$\\
		\hline
	\end{tabular}
	\caption{The estimator $\hat{\theta}$, the determination coefficient ${\cal R}$, the variance $\delta$ and  the convergence rate $\hat{q}$ for LRTO approach by using the initial series of $\tilde{R}'_n(M_\tau)$ up to  $n_\text{th}$-order QCD corrections, where $n=(2,3,4)$, respectively. $\mu_r=M_\tau$. }
	\label{t4}
\end{table}

Third, we apply the LRTO approach to $\tilde{R}_n(M_\tau)$ via an order-by-order way to estimate the UHO-contributions, where $n=(2,3,4)$, respectively. Here similar to any other approaches suggested in the literature, we at least need to know the first two terms to make the LRTO be workable, and we have implicitly set $n\geq2$ to do the estimation. To be consistent with the formulas given in Sec.~\ref{secIII}, the trivial tree-level term ``$1$'' should be kept in the perturbative series, e.g. $\tilde{R}_n(M_\tau)\to \tilde{R}'_n(M_\tau)=1+\tilde{R}_n(M_\tau)$ with $C_0=1$. The parameters such as the estimator $\hat{\theta}$, the determination coefficient ${\cal R}$, the variance $\delta$ of the LRTO approach are put in Tables \ref{t3} and \ref{t4}, which are for PMCs series and initial pQCD series with $\mu_r=M_\tau$, respectively. Tables \ref{t3} and \ref{t4} show that the magnitudes of ${\cal R}$ are consistent with the requirement of the $95.5\%$ significance level, thus the LRTO approach is applicable for both the initial pQCD series and the PMCs series. Such applicability strengths when more loop terms are known, which is also confirmed by the decrement of the variance $\delta$ with the increasing orders. It is known that for a series up to $n_{\rm th}$-order QCD corrections, the convergence rate satisfies $q_{n}\sim \exp(\theta_{n})$. Generally, a smaller convergence rate of the series indicates a better convergent behavior. Numerically, for $\tilde{R}'_n(M_\tau)$, we have $q_{2}$, $q_{3}$, $q_{4}$=$0.2156$, $0.2090$, $0.2221$ for the PMCs series, and $0.1936$, $0.2584$, $0.3000$ for the initial series with $\mu_r=M_\tau$, respectively. A slightly smaller convergence rate of the PMCs series indicates a better convergent behavior than that of the initial series, which is consistent with the decrement of expansion coefficients at higher orders as shown by Table \ref{taucoefficient}.

\begin{figure}[htb]
	\centering
	\includegraphics[width=0.48\textwidth]{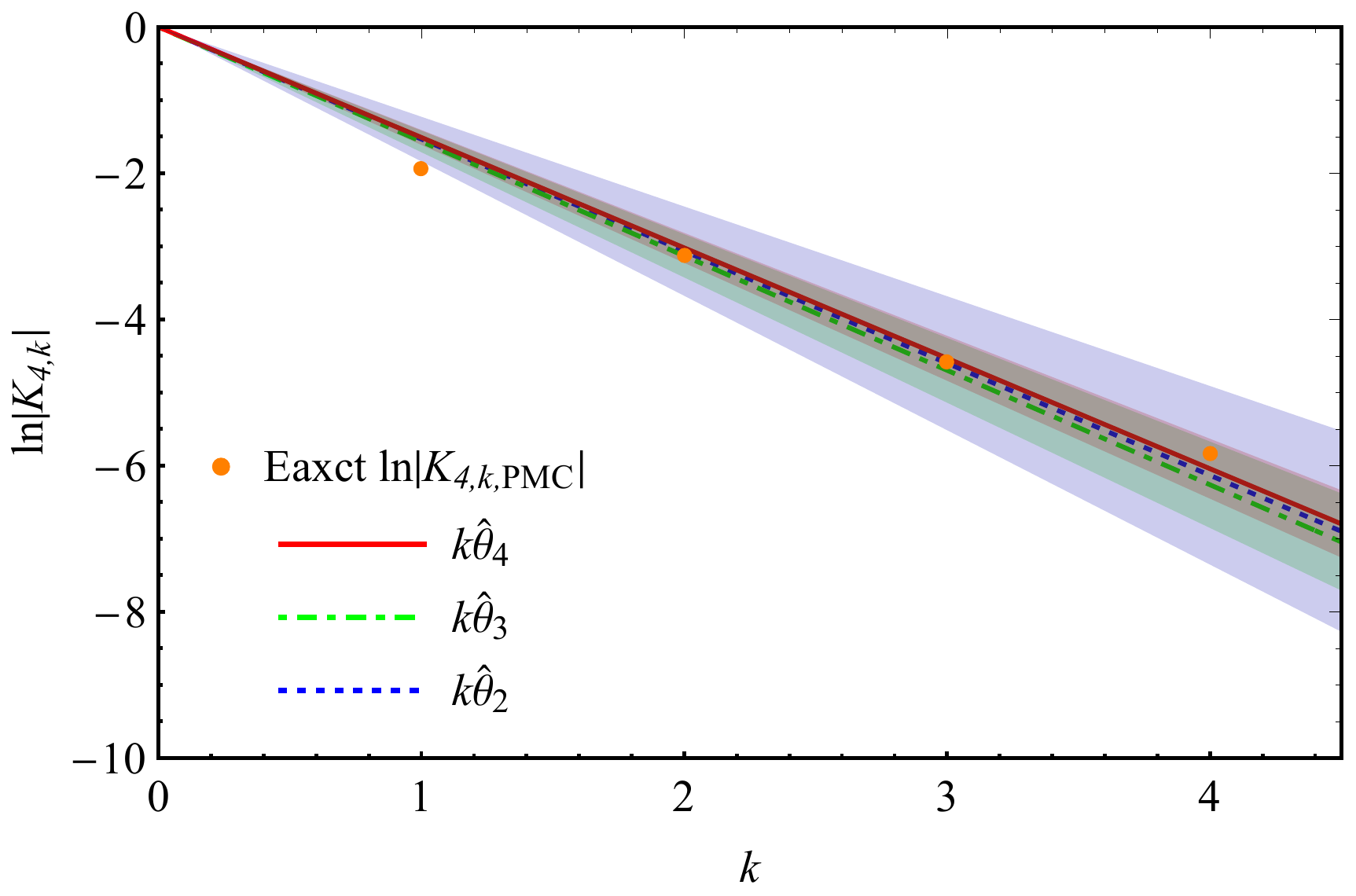} %
	\includegraphics[width=0.48\textwidth]{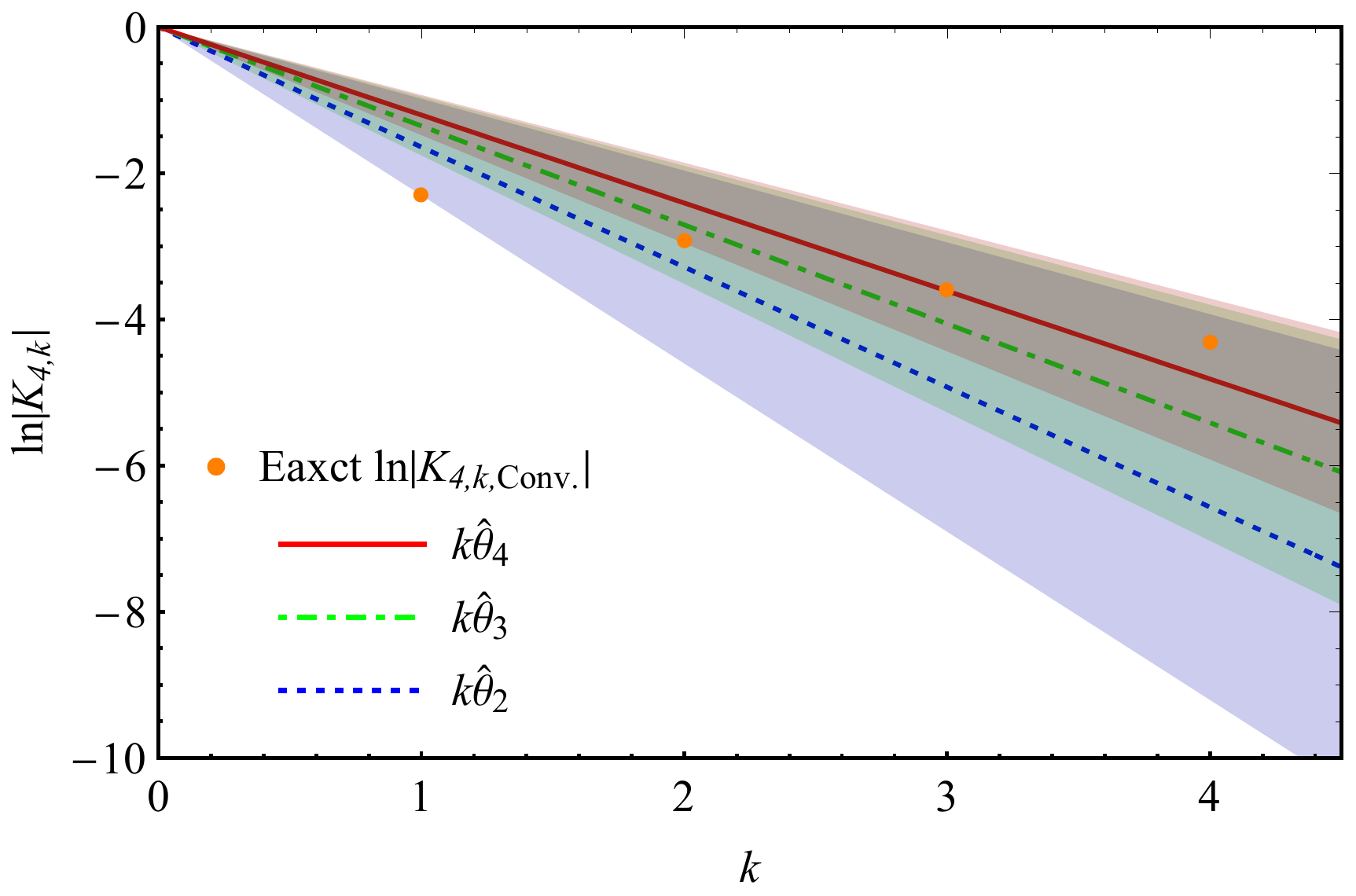} %
	\caption{The LRTO fittings of $\ln |K_{k}|$ before and after applying the PMCs to $\tilde{R}'_{n}(M_\tau)$ up to $n_{\rm th}$-loop QCD corrections, where $k=(2,\cdots,n)$. The solid lines are for $\ln |K_{k}|=k\cdot\hat{\theta}_{k}$, and the shaded bands are the CIs $(2k\cdot\delta_n)$ with DoB=$95.5\%$ as shown in \eqref{CItheta}. The dots are ``exact values'' of $\ln |K_{k}|$ which are calculated by using the known pQCD series before and after applying the PMCs, respectively. }
	\label{Fitting plot}
\end{figure}

We then put the LRTO fittings of $\ln |K_{k}|$ before and after applying the PMCs to $\tilde{R}'_{n}(M_\tau)$ with up to $n_{\rm th}$-loop QCD corrections in Figure~\ref{Fitting plot}, where $k=(2,\cdots,n)$, respectively. Here the solid lines are for $\ln |K_{k}|=k\cdot\theta_{k}$, and the shaded bands represent the CIs $(u_{p/2}k\cdot\delta_4)$ with DoB=$95.5\%$ as shown in \eqref{CItheta} for  $\tilde{R}'_{n}(M_\tau)$. The dots are ``exact values (ECs)'' of $\ln |K_{n,k}|$ which are calculated by using the known pQCD series before and after applying the PMCs, respectively. Figure~\ref{Fitting plot} shows the ``exact value'' of $K$-factors at the orders $n=(2, 3, 4)$ fall well within the credible intervals under the case of DoB=$95.5\%$ $\tilde{R}'_{n}(M_\tau)$. These findings suggest that the power-law behavior of the $K$-factors concerning the order becomes progressively evident with increasing order. 

\begin{figure}[htb]
	\centering
	\includegraphics[width=0.48\textwidth]{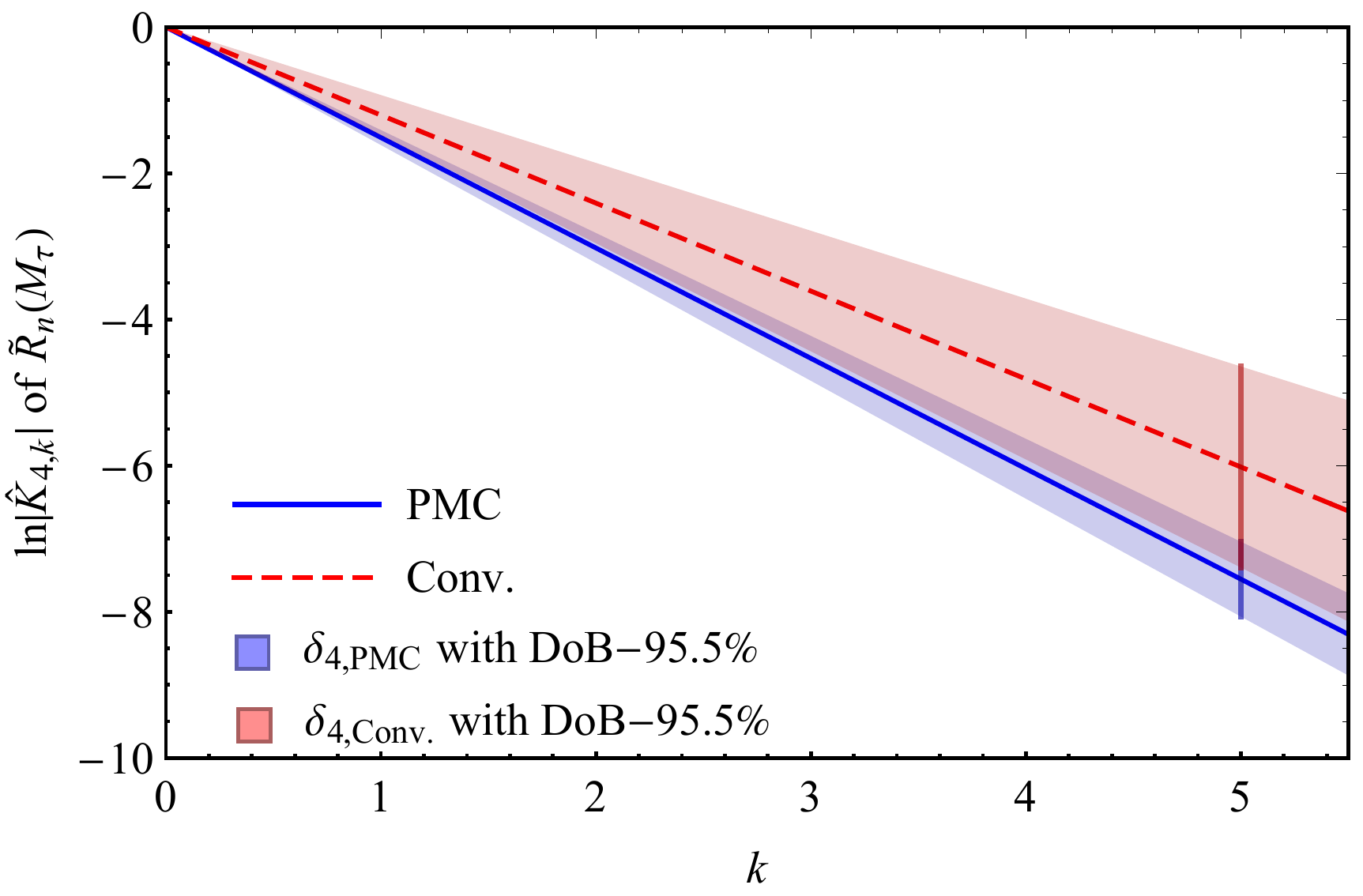} %
	\caption{A comparison of the predicted $\ln |K_{5}|$ by using the LRTO approach for conventional (Conv.) and PMCs series of $\tilde{R}'_{4}(M_\tau)$. The central lines are for $\ln |K_{k}|\simeq k\cdot\theta_{4}$ and the shaded bands represent their CIs $(2k\cdot\delta_4)$ with DoB=$95.5\%$, the red and blue vertical lines for $k = 5$ correspond to the estimation intervals of $\ln|K_5|$ in each of two series with PMCs and Conv. as shown in \eqref{CItheta}. }
	\label{loop4}
\end{figure}

Figure~\ref{Fitting plot} shows more convergent PMCs series does give more precise fittings: 1) the central values are closer to ``exact values'' and more quickly tends to steady value with the increment of loop numbers, e.g. the two-loop, the three-loop and the four-loop predictions are closer to each other for the PMCs series; 2) the error bands are much more smaller for the PMCs series under the same DoB. The one-order higher UHO-contribution is then estimated to be $\ln |K_{n+1,k}|=k\cdot\theta_{n}$. Figure \ref{loop4} shows a comparison of the predicted values of $\ln |K_{5}|$ obtained using the LRTO approach -- both before and after applying the PMCs -- with the four-loop $\tilde{R}'_{4}(M_\tau)$, where $\ln |K_{5}|=5\cdot\theta_{5}$ with the approximation $\theta_5\simeq\theta_4$, and the shaded bands represent their CIs $2k\cdot\delta_4$ with DoB=$95.5\%$ as shown in \eqref{CItheta}, which are calculated with the help of Eqs.(\ref{uho}, \ref{CIrho}). As a foundational result derived from the LRTO approach, it allows us to establish confidence intervals for the parameters of the perturbation series at a specified confidence level $p\%$. While it is more convenient to directly obtain the $K$-factor and the overall contribution for the LRTO model, for the sake of general research practices and gaining a more intuitive understanding, we still start with the coefficients.

\begin{table*}[htb]
	\centering
	\begin{tabular}{c|cccc}
		\hline
		$p\%=95.5\%$ & $|\hat{r}_{2,0}|$ & $|\hat{r}_{3,0}|$ & $|\hat{r}_{4,0}|$ & $|\hat{r}_{5,0}|$\\
		\hline
		CI&0.1013&$\err{0.08062}{0.12093}{0.04837}$&$\err{0.04574}{0.03684}{0.02041}$&$\err{0.02805}{0.01876}{0.01124}$\\
		EC&0.2174&0.11081&0.06982&-\\
		\hline
		$p\%=95.5\%$&$|C_2(\mu_r=M_\tau)|$&$|C_3(\mu_r=M_\tau)|$&$|C_4(\mu_r=M_\tau)|$&$|C_5(\mu_r=M_\tau)|$\\
		\hline
		CI&0.1013&$\err{0.2333}{1.4549}{0.2011}$&$\err{0.4352}{1.753}{0.3486}$&$\err{0.7349}{2.1740}{0.5492}$\\
		EC&0.5271&0.8503&1.3055&-\\
		\hline
	\end{tabular}
	\caption{The predicted CIs for scale-independent conformal coefficients $\hat{r}_{i,0}(i=2,3,4,5)$ and scale-dependent initial coefficients $C_i(\mu_r)(i=2,3,4,5)$ at the scale $\mu_r=M_\tau$ of $\tilde{R}_n(M_\tau)$ via the LRTO. The exact values (``ECs''), which are calculated by using the known pQCD series under the same input parameters, are presented for comparison. }
	\label{t7}
\end{table*}

\begin{figure}[htb]
	\centering
	\includegraphics[width=0.48\textwidth]{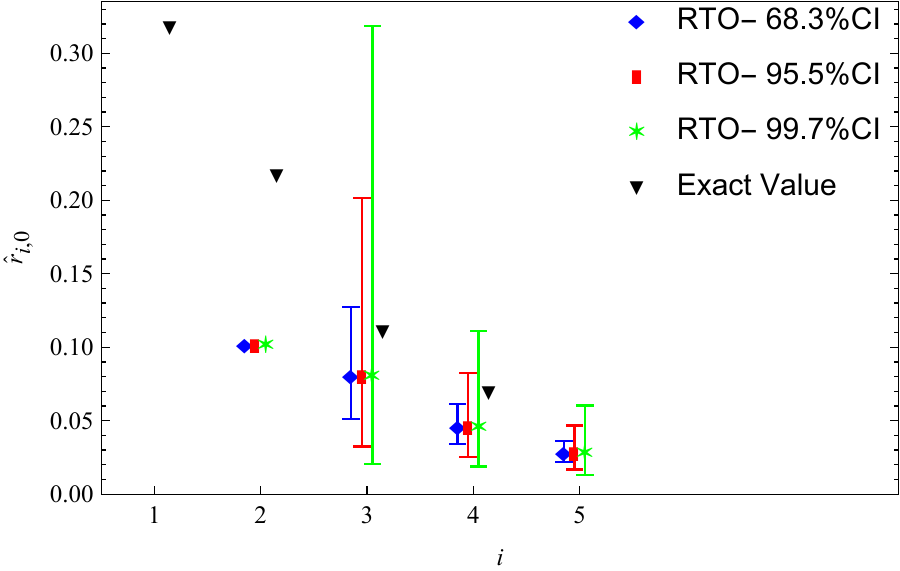}
	\includegraphics[width=0.48\textwidth]{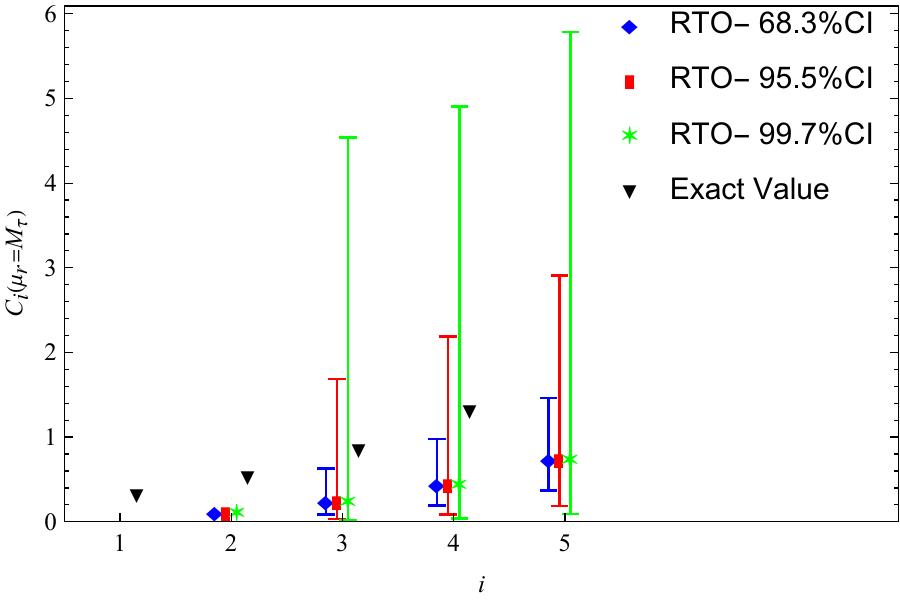}
	\caption{The predicted CIs with three DoBs for the scale-independent conformal coefficients $\hat{r}_{i,0}$ and the scale-dependent initial coefficients $C_i(\mu_r)$ at the $\mu_r=M_\tau$ under the LRTO for $\tilde{R}'_n(M_\tau)$. The blue diamonds, the red rectangles, the green stars and the black inverted triangles together with their error bars, are for 99.7\% CI, 95.5\% CI, 68.3\% CI, and the exact values (``ECs'') of the coefficients at different orders, respectively. The exact values, which are calculated by using the known pQCD series under the same input parameters, are presented for comparison.}
	\label{cplot}
\end{figure}

Fourth, we provide the $95.5\%$ confidence interval (CI) and the precise values of the scale-independent conformal coefficients $|r_{i,0}|$ (for $i=2,3,4,5$) from the PMC series of $\tilde{R}_n(M_\tau)$ in Table \ref{taucoefficient}. For comparison, we also present the similarly predicted scale-dependent conventional coefficients $|C_i(\mu_r)|$ (for $i=2,3,4,5$) of the conventional series of $R_n(Q=\text{31.6GeV})$ at the scale $\mu_r=Q$ and $\tilde{R}_n(M_\tau)$ at the specific scale $\mu_r=M_\tau$, where the results for 99.7\% CI, 95.5\% CI, 68.3\% CI are presented accordingly. It's worth noting that the exact values of $|\hat{r}_{3,0}|, |\hat{r}_{4,0}|, |C_3|$, and $|C_4|$ fall well within the 95.5\% CI for $\tilde{R}_n(M_\tau)$. Figure~\ref{cplot} also 
shows that when the coefficients all exhibit a consistent trend -- e.g., decreasing -- the LRTO method can estimate these coefficients effectively, with smaller errors.

\begin{figure*}[htbp]
	\centering
	\includegraphics[width=0.46\textwidth]{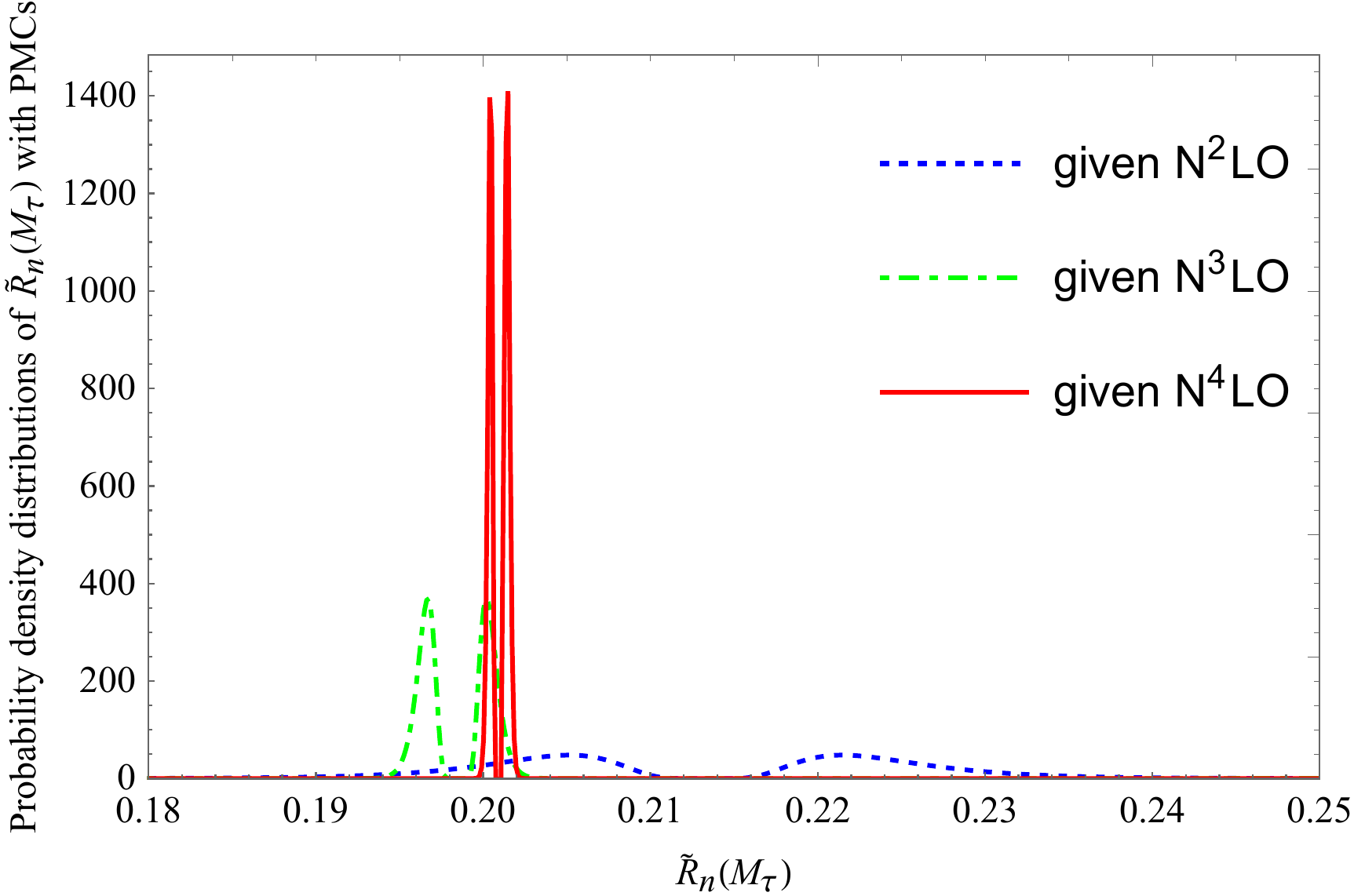}
	\includegraphics[width=0.46\textwidth]{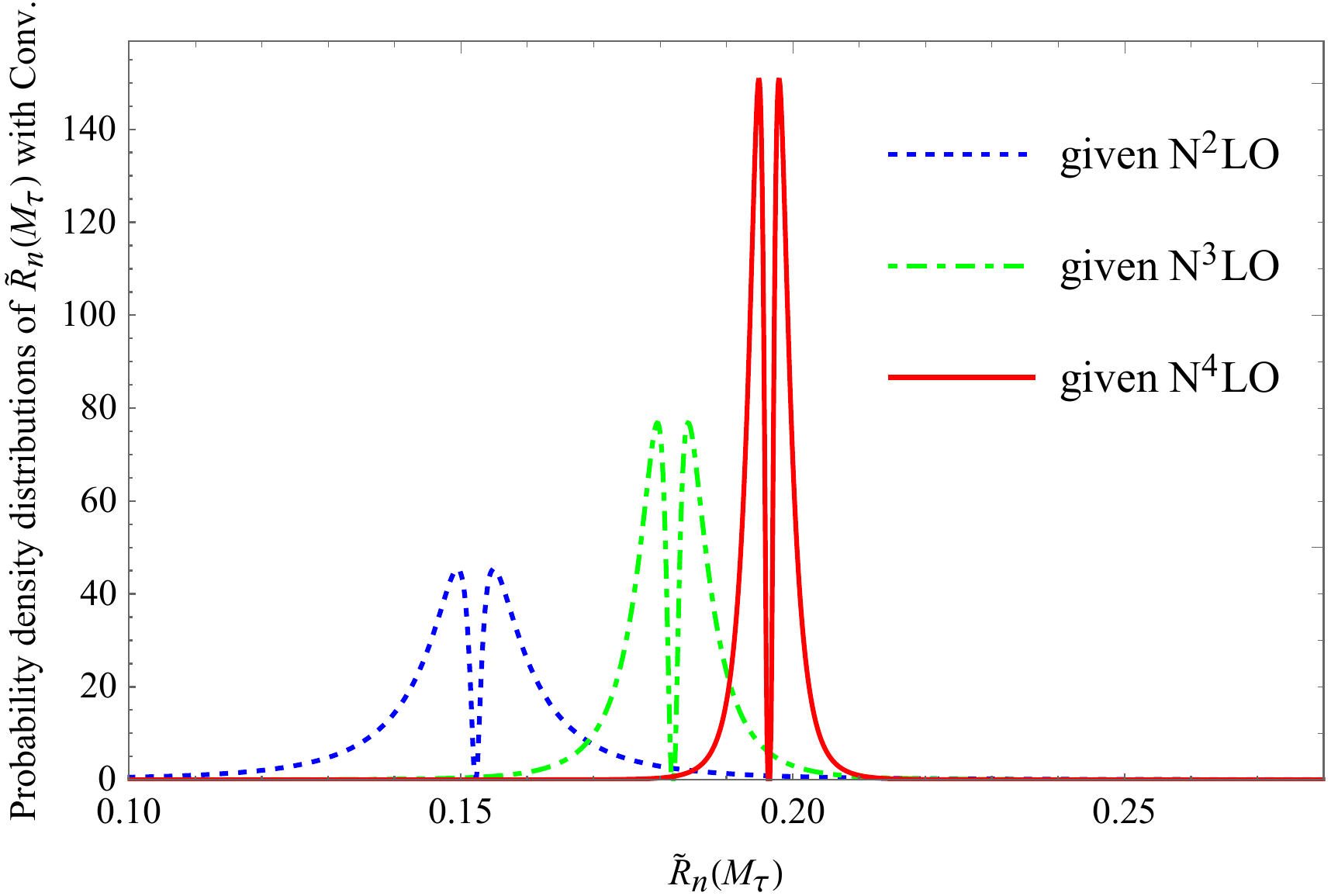}
	\caption{The probability density distributions of $\tilde{R}(M_\tau)$ with PMCs and conventional (Conv.) series at different states of knowledge predicted by the LRTO, respectively. The dashed, the dash-dot and the solid lines are results for the given $\mathrm{N^2LO}$, $\mathrm{N^3LO}$ and $\mathrm{N^4LO}$ series, respectively.}
	\label{pdfplot}
\end{figure*}

Fifth, we present the probability density distributions for $\tilde{R}(M_\tau)$ with PMCs and conventional series at different states of knowledge predicted by the LRTO in Figure~\ref{pdfplot}. The three lines correspond to different degrees of knowledge: the given $\mathrm{N^2}$LO (dashed), the given $\mathrm{N^3}$LO (dash-dot) and the given $\mathrm{N^4}$LO (solid), respectively. Figure~\ref{pdfplot} illustrates the characteristics of the log-normal distribution, a skewed distribution with values greater than (the other half of the figure is symmetric due to the absolute value, and the symmetric center represents the calculated value of this order). Eq.\eqref{LGD} shows that the spacing of the twin peaks indicates the magnitude of the UHO-terms, and the sharpness of the peaks reflects the uncertainties introduced by fluctuations in the coefficient. 

\begin{table*}[htb]
	\centering
	\begin{tabular}{c|c|cc|cc|c}
		\hline
		$p\%=95.5\%$ & EC, $n=2$ & CI, $n=3$ & EC, $n=3$ & CI, $n=4$ & EC, $n=4$ & CI, $n=5$\\
		\hline
		$\tilde{R}_n|_\mathrm{PMCs}$&0.2133&$[0.1882,0.2385]$&0.1985&$[0.1950,0.2019]$&0.2009&$[0.2001,0.2018]$\\
		$\tilde{R}_n(M_\tau)|_\mathrm{Conv.}$&0.1522&$[0.0997,0.2047]$&0.1819&[0.1595,0.2043]&0.1964&[0.1868,0.2061]\\
		\hline
	\end{tabular}
	\caption{ Comparison of exact values (``ECs'') and the predicted given DoB CIs of the pQCD approximants, $\tilde{R}_n$ calculated by using the PMC series and the conventional (Conv.) scale-dependent series up to $n_\mathrm{th}$-order level, respectively. The exact values are calculated by using the known series under the same input parameters. The results for the PMC series are scale-independent. The results for the conventional series are calculated by fixing $\mu_r=M_\tau$ for $\tilde{R}_n(M_\tau)$.}
	\label{t6}
\end{table*}

\begin{figure}[htb]
	\centering
	\includegraphics[width=0.48\textwidth]{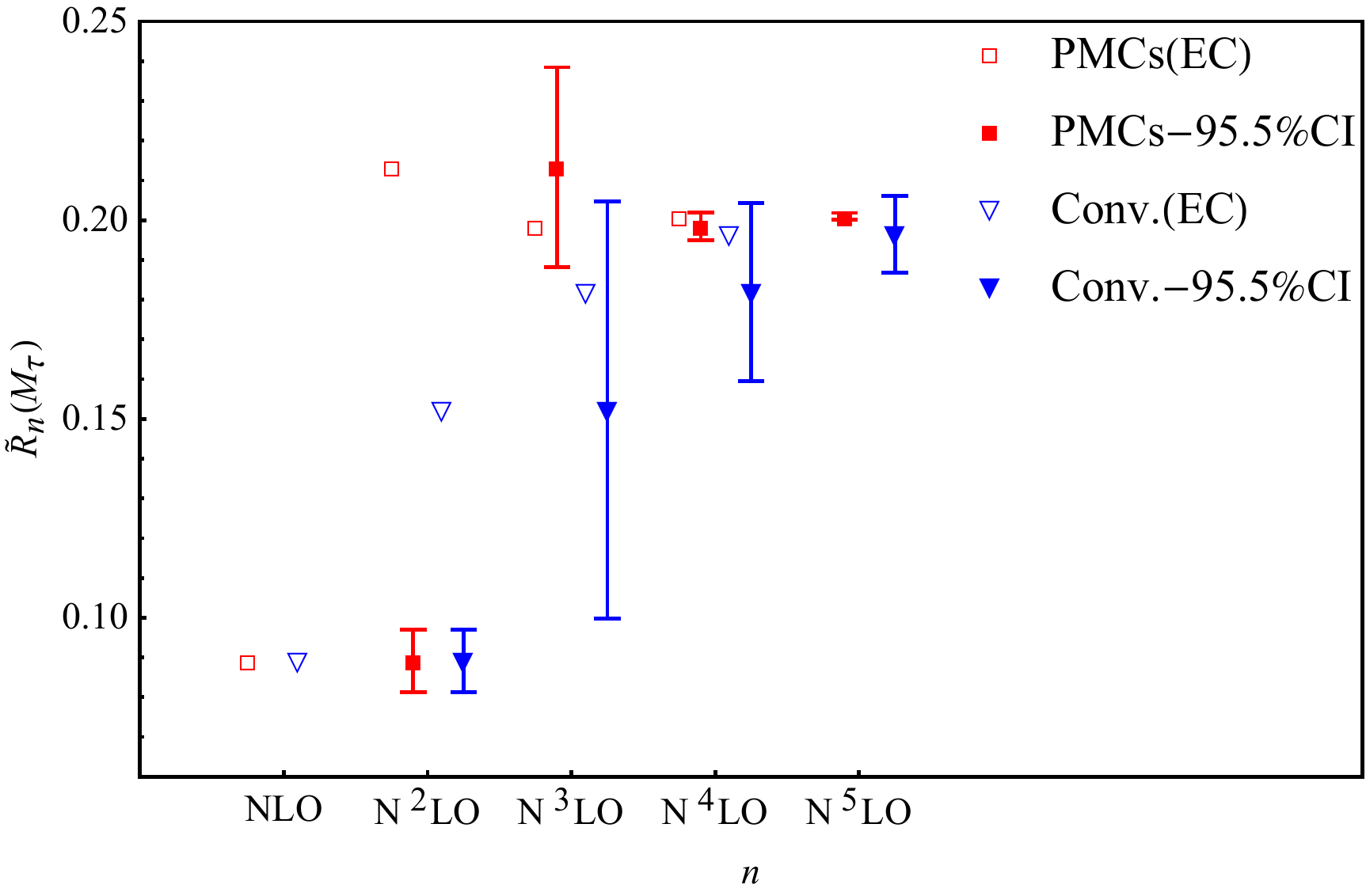}
	\caption{Comparison of the calculated central values (the pQCD approximants $\tilde{R}_n(M_\tau)$ with the predicted 95.5\% CIs of those approximants for $n = (2, 3, 4, 5)$. The blue hollow triangles and the red hollow quadrates represent the calculated central values of the fixed-order pQCD predictions using Conv. and PMCs scale-setting approaches, respectively. The blue solid triangles and the red solid quadrates with error bars represent the predicted 95.5\% CIs for $\tilde{R}_n(M_\tau)$ using the LRTO based on the PMCs conformal series and the conventional (Conv.) scale-dependent series, respectively.}
	\label{f3}
\end{figure}

Sixth, we present $95.5\%$ CIs of $\tilde{R}_n(M_\tau)(n=2,3,4,5)$ from the one order lower $\tilde{R}_{n-1}(M_\tau)$ by the LRTO in Table \ref{t6} and Figure \ref{f3}. As expected, the error bars given by the LRTO gradually decrease with the increasing known orders, which is a reflection of the convergence of the series. The exact values of $\tilde{R}_n$ for both the PMCs series and conventional series. The exact values of $\tilde{R}_{n}$ are calculated by using the known pQCD series under the same input parameters, which are well within the error bars estimated by the LRTO from one order lower given series $\tilde{R}_{n-1}$. Figure \ref{f3} shows that the estimations of the PMCs series show better convergent behavior at any known orders, which are also confirmed by Table \ref{t6}. 

Using LRTO method, our final results for the five-loop results of $\tilde{R}_5$ for both PMCs and conventional scale-setting approaches are
 \begin{align}
	\tilde{R}_5(M_\tau)|_\mathrm{Conv.}&=\err{0.1964}{0.0128}{0.0291}\pm{0.0096}\label{R5Conv.},\\
	\tilde{R}_5(M_\tau)|_\mathrm{PMC}&=\err{0.2009}{0.0054}{0.0082}\pm{0.0009},\label{R5PMCs}
\end{align}
Here the first error of Eq.\eqref{R5Conv.} is caused by variation of scale from $1/2 M_\tau$ to $2M_\tau$ and the first error of Eq.\eqref{R5PMCs} is the influence caused by UHO-term to $Q_*$ which can be estimated by conservative estimation. The $|S_3|=|S_2+|S_2-S_3||$ which leads to a scale shift $\Delta{Q_*}=\pm0.03$ GeV.

\section{summary} \label{secV}

It has been stated that the PMC prediction satisfies the RGI and eliminates the conventional renormalization scale and scheme dependence at any fixed order. The PMC series becomes more convergent due to the elimination of divergent renormalon term. Our present results confirm those PMC features and demonstrate that the PMC single-scale series serves as a solid foundation for the estimation of the UHO contribution. Thereby, the applicability of pQCD theory is enhanced.

We have proposed a novel method for estimating the contributions from the UHO terms, and demonstrated its application with one concrete example. The method, termed LRTO, quantifies the convergence behavior of the perturbative series prior to its optimal truncation. The core of LRTO lies in modeling the dominant exponential trend of the series coefficients, as described by the relation (\ref{kmodel}), which leads to the linear relation (\ref{linearRelation}), thereby transforming the estimation of the UHO-contributions into an estimation of the linear parameter $\theta \equiv \ln q$, which characterizes the convergence rate. In those two equations, the sub-leading corrections $\epsilon_{k}$ are treated as regression residuals, and their variance $\delta$ is incorporated into the theoretical uncertainty of the extracted parameter $\theta$, which also provides the method with a tolerance for data fluctuations and a probabilistic interpretation. Finally, the determination coefficient, quantified by a coefficient such as $\mathcal{R}$, describes the applicability and reliability of LRTO for the specific process under study. Together, the convergence rate parameter $\theta$, its associated uncertainty $\delta$, and the determination coefficient $\mathcal{R}$ form a complete evaluative framework. This tripartite structure and its results are visualized in Figs. \ref{Fitting plot}, \ref{loop4} and summarized in Tables \ref{t3}, \ref{t4}. \footnote{For the case of expansion coefficients exhibiting significant oscillatory behavior, a suitable preprocessing step to mitigate the oscillations is necessary before applying the LRTO, especially if the coupling $\alpha_s$ is not sufficiently small. We note that at higher orders, the underlying convergence trend typically reasserts itself, allowing LRTO to yield robust results. A detailed discussion of this aspect is in preparation.}
	
The results under the PMCs and Conventional scale-setting approaches, using the LRTO, have been carefully analyzed. Tables \ref{t7},\ref{t6}. and Figs.\ref{Fitting plot},\ref{cplot},\ref{pdfplot},\ref{f3} indicate that the LRTO demonstrates its own applicability to both PMCs and conventional pQCD series and achieves estimations that complies with expectations. Using the PMCs approach, a more convergent, stable and reliable series can be obtained. As a comparison of the results of $\tilde{R}_n$ from the perspective of LRTO, the PMCs series shows a much better convergent behavior compared to conventional series, even by fixing the scale of conventional series by $\mu_r=M_\tau$. And with the help of LRTO, the PMCs series are visualized even more. Therefore, based on the above results, by comparing the effect of non-conformal and conformal terms on the series, it can be learned that the non-conformal terms of the series will be the main source of error when the perturbation order increases, which proves the necessity of using a proper scale-setting approach such as PMCs at any fixed orders.

Our results show that the LRTO method can be served as an important method for estimating contributions of the UHO-terms. LRTO gives a way to judge the convergence of a series and estimates UHO-terms by using the convergence rate $q$. PMC series with scale-invariant and more convergence exhibits a much better predictive power with stability and reliability than the initial scale-dependent pQCD series. Thus a combination of those two treatments can be applied to improve the precision and predictive power of the pQCD theory.

\hspace{1cm}

{\bf Acknowledgments:} This work was supported in part by the Chongqing Graduate Research and Innovation Foundation under Grant No.CYB23011 and No.ydstd1912, and by the Natural Science Foundation of China under Grant No.11905056, No.12175025 and No.12347101.

\end{document}